\documentclass[a4paper,fleqn]{cas-sc}
\usepackage[numbers]{natbib}
\usepackage{subfigure}
\usepackage{epstopdf}

\def\tsc#1{\csdef{#1}{\textsc{\lowercase{#1}}\xspace}}
\tsc{WGM}
\tsc{QE}
\tsc{EP}
\tsc{PMS}
\tsc{BEC}
\tsc{DE}

\begin{document}
\let\WriteBookmarks\relax
\def\floatpagepagefraction{1}
\def\textpagefraction{.001}
\shorttitle{Enhanced Residual SwinV2 Transformer for Learned Image Compression}
\let\printorcid\relax

\title [mode = title]{Enhanced Residual SwinV2 Transformer for Learned Image Compression}

\author[1]{Yongqiang Wang}
\cormark[1]
\ead{wangyq0901@163.com}
\author[1]{ Feng Liang}
\author[1]{ Haisheng Fu}
\author[2]{ Jie Liang}
\author[1]{ Haipeng Qin}
\author[2]{ Junzhe Liang}

\address[1]{School of Microelectronics, Xi'an Jiaotong University,  China}
\address[2]{School of Engineering Science, Simon Fraser University, Canada}

\begin{abstract}
Recently, the deep learning technology has been successfully applied in the field of image compression, leading to superior rate-distortion performance. However, a challenge of many learning-based approaches is that they often achieve better performance via sacrificing complexity, which making practical deployment difficult. To alleviate this issue, in this paper, we propose an effective and efficient learned image compression framework based on an enhanced residual Swinv2 transformer. To enhance the nonlinear representation of images in our framework, we use a feature enhancement module that consists of three consecutive convolutional layers. In the subsequent coding and hyper coding steps, we utilize a SwinV2 transformer-based attention mechanism to process the input image. The SwinV2 model can help to reduce model complexity while maintaining high performance. Experimental results show that the proposed method achieves comparable performance compared to some recent learned image compression methods on Kodak and Tecnick datasets, and outperforms some traditional codecs including VVC. In particular, our method achieves comparable results while reducing model complexity by 56\% compared to these recent methods.
\end{abstract}


\maketitle

\section{Introduction}

Recently, the application of deep learning to image compression has gradually outperformed traditional approaches. The main purpose of image compression is to reduce space redundancy for transmission and storage. Some traditional compression standards such as JEPG \cite{wallace1992jpeg}, JEPG2000 \cite{taubman2002jpeg2000}, Better Portable Graphics (BPG) \cite{Bpg_125} and Versatile Video Coding (VVC) \cite{H266} can effectivately improve compression performance via linear transforms such as the discrete cosine transform (DCT) \cite{ahmed1974discrete} and discrete wavelet transform (DWT) \cite{wavelet}. However, the handcrafted transforms will cause block effects and ringing blurry artifacts \cite{ahmed1974discrete}. Similar to traditional codecs, the learning-based image compression framework also includes transform, quantization, and entropy coding. Each module is composed of a learnable network in learning-based image compression architectures.  

Most existing the learning-based image compression networks are based on Variational Autoencoder (VAE) architecture \cite{ICLR2018_26}. The VAE-based image compression methods could capture the underlying distribution of features in the original data during encoding, and then use it to generate similar data during decoding. Most methods improve upon this architecture, including those known as generalized divisive normalization (GDN) \cite{GDN} and non-local attention module \cite{cheng2020learned}. After data transformation, quantization operations are performed on the floating point outputs of the network. However, quantization operations are not differentiable and need to be approximated using some alternative methods. One widely used approach is additive uniform noise, as proposed in \cite{ICLR2017_67}. In \cite{NIPS2017_6714}, the soft-to-hard vector quantization is utilized to replace the round quantization in \cite{8683541} \cite{fu2020improved}.

In order to accurately estimate the probability distribution of the latent representations, it is crucial to design an efficient entropy model. Previous works have made significant efforts to address this challenge. For example, in \cite{ICLR2018_26},  a scale hyperprior based on a single Gaussian model is propose, in which the scale parameters are estimated by a hyperprior.  Based on \cite{ICLR2018_26}, Cheng et al \cite{cheng2020learned} have made further strides in improving the scale hyperprior by incorporating attention modules and discretized Gaussian mixture likelihoods to better parameterize latent features, leading to  significant improvements in compression efficiency. However, the previous works only use the single distribution, the latent representations still exist some spatial redundancy. To solve these problems, the Gussian mixture Gaussian-Laplacian-Logistic Mixture Model (GLLMM) is proposed in \cite{fu2021learned}.  

Many VAE-based encoding architectures stack multiple convolutional layers to extract local spatial correlation information. However, they often struggle to capture long-distance features, which leads to underutilization of important information. To further extract global information, some image compression models based on transformer are proposed. Zuo et al.\cite{Zou_Song_Zhang} propose a window-based attention to capture the spatial neighboring elements correlations. In \cite{qian2022entroformer}, the author propose entroformer model based on ViT \cite{vit} model, which enables joint learning of both spatial and content information. Additionally, they expend the bidirectional parallel context model, resulting in faster decoding process. Lu et al. \cite{lu2021transformer} propose a method based on Swin transformer to acquire short-range and long-range information learning of images. A recently proposed in \cite{Liu_Sun_Katto_2023} parallel transformer-CNN model realizes the parallel combination of CNN's local modeling capability and transformer's non-local modeling capability to achieve state-of-the-art performance.
In this paper, we mainly proposes the Enhanced Residual SwinV2 Transformer for learned image compression method. We combine a convolutional layer and the Residual SwinV2 Transformer Block (RS2TB) to characterize the spatial information. Different from the Swin transformer, residual Swinv2 block was used to help the model train stably through the operation of post-norm and cosine similarity. At the same time, the model parameters are further reduced. Similarly to \cite{lu2021transformer}, we also utils a causal attention module (CAM) to encode the hyper priors. Additionally, we introduce a feature enhancement module before the RS2TB to improve the non-linear representativeness of our network. Specifically, this module is based on the popular Dense Block \cite{huang2017densely}.
In summary, the contributions of this paper can be summarized as follows:
\begin{itemize}
\item Inspired by \cite{lu2021transformer}, we develop the SwinV2 transformer for image Compression method. Different from Swin transformer, post-normalization technique and cosine attention are used in SwinV2 transformer, which can greatly improve model stability.

\item To enhance the non-linear representational capacity of our network, we have incorporated a feature enhancement module \cite{xie2021enhanced} in a residual manner before the RS2TB architecture. This module is based on the widely used Dense Block \cite{huang2017densely}, and is composed of three consecutive convolutional layers with kernel size 1, 3, and 1, respectively.

\item Compared with the complexity of other recent method \cite{cheng2020learned}, the proposed method can save nearly half of the model size under the same compression efficiency. The coding time is similar at low bit rates, but our model requires only 57.09\% of the model size. At high bit rates, our coding time is significantly less than but the model size is only 56.81 \% of them.

\end{itemize}

Thanks for these contributions, experimental results using the Kodak\cite{Kodak_dataset} and Tecnick  \cite{Tecnick_dataset} datasets show that the proposed scheme outperforms some recent works in terms of PSNR and MS-SSIM. Compared to Cheng\cite{cheng2020learned}, our schemes achieve better performance on PSNR at high bit rate, especially when the bpp is greater than 0.5. At a compression ratio of 0.8bpp, our model achieves a PSNR improvement of 0.23dB, which is almost comparable to the performance of VVC. Additionally, when optimizing for MS-SSIM on the Tecnick dataset, our method outperforms some recent learned models.

\begin{figure*}
	\centering
		\includegraphics[scale=0.6]{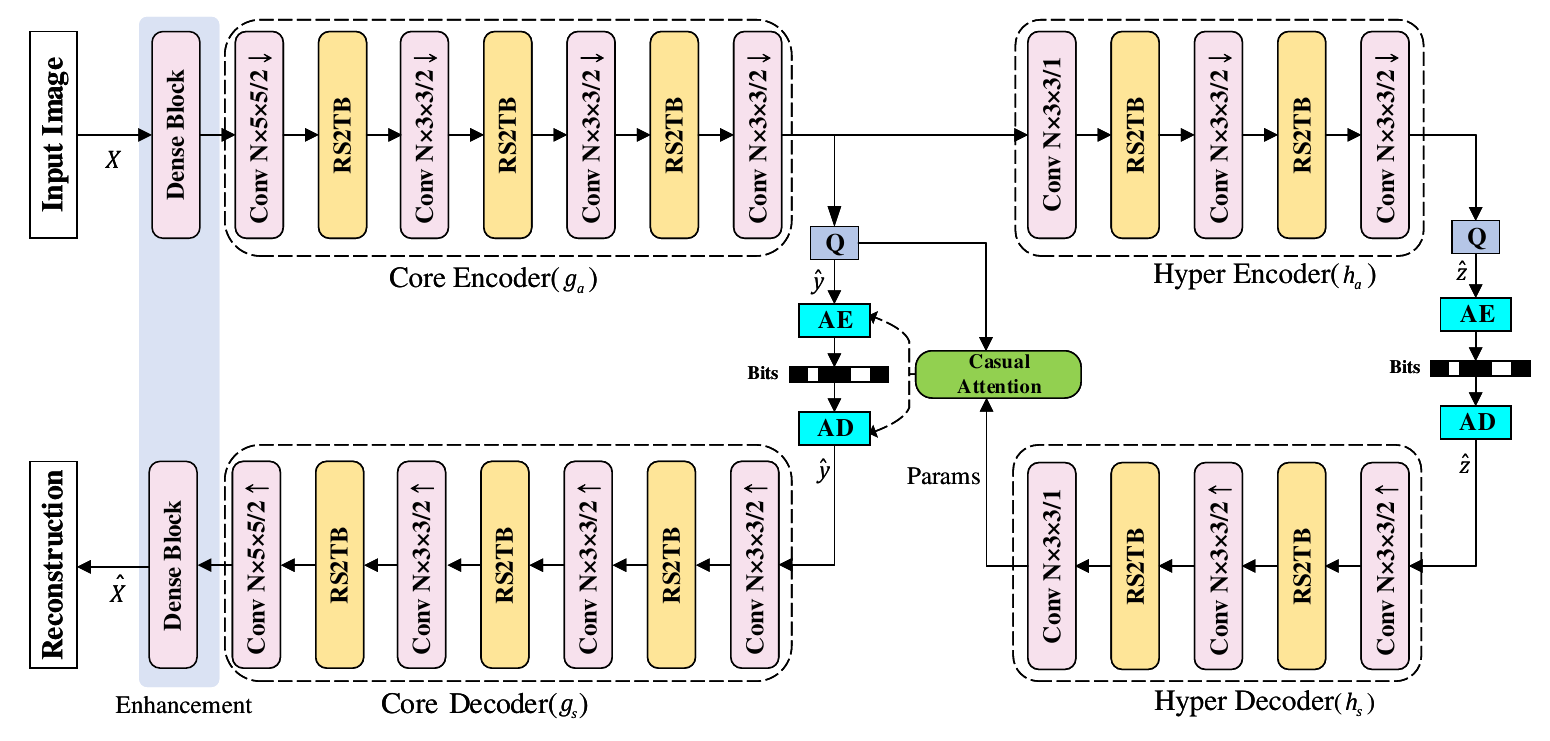}
	\caption{The detailed structures of the proposed image compression framework. $\uparrow$ and $\downarrow$ represent the up- or down- sampling operation. $5\times5$ and $3\times3$ represent the convolution kernel size. AE and AD  stand for arithmetic encoder and arithmetic decoder, respectively.}
 
	\label{FIG:1}
\end{figure*}

\begin{figure*}
\centering
\subfigure[RS2TB]{
\begin{minipage}[t]{0.33\linewidth}
\centering
\includegraphics[scale=0.60]{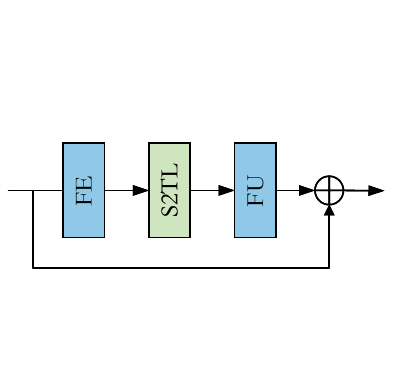}
\end{minipage}
}%
\subfigure[SwinV2 Transformer]{
\begin{minipage}[t]{0.33\linewidth}
\centering
\includegraphics[scale=0.60]{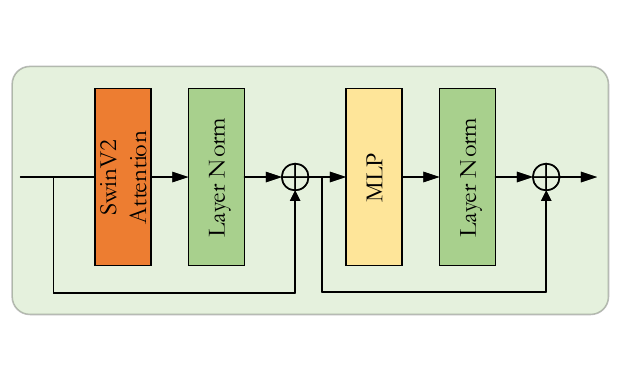}
\end{minipage}
}%
\subfigure[SwinV2 Attention Module]{
\begin{minipage}[t]{0.33\linewidth}
\centering
\includegraphics[scale=0.60]{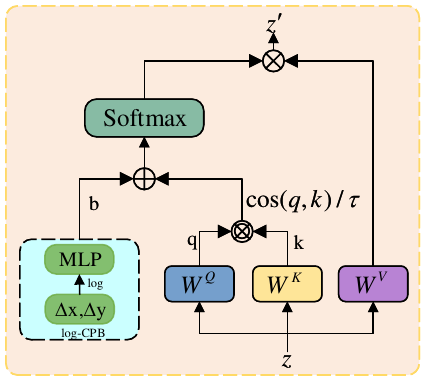}
\end{minipage}
}%
\centering
\caption{The detailed structures of the RS2TB. (a)The Residual SwinV2 Transformer Block. (b)The SwinV2 transformer block. (c)The SwinV2 Attention Module.}
\label{FIG:2}
\end{figure*}

\section{Related Work}
\subsection{Traditional Image Compression Codecs}

Most existing traditional image compression standards adopt manual methods. Specifically, transform, quantization and entropy coding are all designed to remove different types of redundancies. However, these method still have certain limitations. First, in the traditional compression codecs, since the input image is divided into image blocks, there will be block effects after transform and quantization in this way. Second, the reconstructed images of these traditional image compression (such as JPEG, JPEG200 and Webp) will suffer from blurring and ringing artifacts at low bit rates.

The traditional image compression standards mainly use block-based hybrid coding methods. With the development of different coding standards, the block division structure has evolved from a consistent division structure to a separable division, which can efficiently adapt to the encoding and decoding processing of high-resolution images. However, it is crucial to balance the trade-off between encoding performance and complexity.

\subsection{Learned Image Compression Methods}
Image compression is mainly divided into lossless compression and lossy compression. Most learned image compression methods belong to lossy compression. In \cite{johnston2018improved},  the learning-based image compression method based on a recurrent neural network is proposed. The convolutional LSTM is used to achieve a variable-rate image compression framework. In \cite{toderici2015variable}, a spatially adaptive bit rate is proposed.

Recently, the variational autoencoder (VAE) architecture has  been widely used in the field of image compression. To address the non-differentiable problem after quantization, the GDN is proposed to achieve an end-to-end image compression framework. Later in \cite{ICLR2018_26}, the author proposed a super prior model and used GDN for local gain. The CNN compression framework lays the groundwork.
In \cite{cheng2020learned} \cite{fu2021learned}, they changed the single Gaussian probability model into a mixed Gaussian model, which is much more flexible and accurate in estimating the probability distributions of the latent representations.

\subsection{Swin Transformer}
Lately, due to its excellent global feature extraction ability, transformers have achieved significant results in computer vision tasks\cite{vaswani2017attention}. Recent works introduces vision transformer(ViT) has been successfully applied to the computer vision tasks\cite{crossvit,zhou2021deepvit,han2021transformer,wu2021cvt}. In \cite{bai2022towards}, the authors propose an end-to-end image compression and analysis model with transformers. Aiming at the global information redundancy in image compression, \cite{qian2022entroformer} proposes a transformer-based probability model to predict potential features. A Transformer based Image Compression (TIC) \cite{lu2021transformer} approach is developed which reuses the canonical VAE architecture with paired main and hyper encoder, which is based on Swin transformer \cite{liu2021swin}. Besides, a casual attention module (CAM) is devised for adaptive context modeling of latent features to utilize both hyper and autoregressive priors. In \cite{ROI_swin_trans}, a Region Of Interest (ROI) mask based on Swin transformer block is integrated into the compressed network to provide spatial feature, which achieves higher ROI PSNR.

In SwinV2 \cite{liu2022swin}, in order to further scale the capacity of the model and the resolution of the window, the window self-attention module is mainly modified. The original Swin transformer utilizes prenormalization, which merges the output activation value of each residual module with that of the main branch. However, this will caused instability during training, as the amplitude of the main branch increased with each deeper layer. In order to effectively solve this problem, post-normalization is used in SwinV2. The output of each residual module is normalized first and then merged with the main branch, so that the amplitude of the main branch will not be accumulated layer by layer. In the original self-attention calculation, the pixelation of pixel pairs is calculated by the dot product of query and key, but in the large model, the attention map of some modules and head is dominated by a small number of pixel pairs. To alleviate this issue, the Scaled Cosine Attention(SCA) is used, the main equation is shown as follows:
 \begin{equation}\label{Swinv2Attention}
 \text{Attention}(Q, K, V) = \text{Softmax}\left(\frac{\text{cos}(Q, K)}{\tau} + S\right) V
 \end{equation} 
where Q, K, V are the query, key and value matrices, respectively. S are the relative to absolute positional embeddings obtained by projecting the position bias after re-indexing. $\tau$is a learnable scalar, non-shared across heads and layers. This block is illustrated in Fig. \ref{FIG:2}. Finally, a log space continuous position bias method is introduced to make the relative position bias smooth across the window resolution.

In this paper, we attempt to propose an effective and efficient compress framework. we update the Swin transformer block (STB) in \cite{lu2021transformer} by using the new SwinV2 transformer \cite{liu2022swin}, which not only can extract global information, but also  half the model parameters at the same performance compared to other state-of-the-art models.

\section{The proposed Image Compression Framework}

The proposed network architecture is illustrated in Fig. \ref{FIG:1}. The input image has a size of $W\times H\times 3$, where W, H, and 3 represent the length, width, and channel of the input image, respectively. The architecture consists of three sub-networks: feature enhancement, core subnetworks, and hyper subnetworks. To further enhance compression performance,  we incorporate a feature enhancement network using a Dense Block architecture to enhance the non-linear representation of input images. Different from the method in \cite{xie2021enhanced}, we only utilize a single-level Dense Block instead of using a residual block connection. The proposed method could significantly reduce the model parameters and improve the efficiency of training during the training process. The effect is similar to previous methods \cite{cheng2020learned} \cite{chen2021}, but our model requires fewer parameters then theirs.

In the core encoder network, we transform the input image x into the latent representation y. The latent representation is quantized $\hat{y}$, and the entropy network is used to learn the probabilistic model of the quantized latent representation. And then the quantized $\hat{y}$ is encoded to the bitstream via entropy coding. After the input image passes through the nonlinear feature enhancement module, a $5 \times 5$ convolution downsampling operation is firstly used to reduce the calculation amount in the transform and expand the receptive field. The data is then fed into an analysis transform $g_{a}$ containing a three-level transformation to obtain the latent representation y. At each level of transformation, a RS2TB and a Conv3x3 downsampling are included to extract the relevant information. In the hyper encoder network, a similar processing architecture is employed, but we only use two-level transformation modules. The specific architectural information is described in the next section. Although transformer-based image compression is already used in TIC \cite{lu2021transformer}, SwinV2 is adopted in our scheme to improve the stability of the model with the use of post-normalized technology and cosine attention.  

A causal attention module is proposed in \cite{lu2021transformer}, in which CAM expands the quantized features into $5 \times 5$ chunks. This module also uses the masked attention to calculate the relationship between these blocks (MA) to ensure causality.  In this paper, we apply CAM to our model to select attention neighbors from autoregressive prior. Then, the attention-weighted autoregressive neighbors and the superpriors from the hyper decoder $h_{s}$ are integrated in the MLP layer for final context prediction.

\subsection{Enhancement Module}
The Dense Block (DB) is proposed in \cite{huang2017densely}, which improves the flow of information in the network via using a fully connected mode. Different from the traditional convolutional neural networks whose output of each layer is only connected to the input of subsequent layers, DB allows the output of each layer to be connected with the input of all subsequent layers. This results in feature reuse and strengthens the transmission and fusion of features.  In \cite{xie2021enhanced}, a feature enhancement module is added in a residual manner to improve the non-linear representation. 

In this paper, different from \cite{xie2021enhanced} which uses multiple Dense Blocks, we use only one Dense Block for feature enhancement to extract image information . Our method could significantly reduce the number of learned parameters of the network, the model complexity and training time.

\subsection{Residual SwinV2 Transformers Block}
The SwinV2 architecture introduces modifications to the shifted window self-attention module to enable better scaling of model capacity and window resolution. By utilizing post normalization, the average feature variance of deeper layers is reduced, leading to increased numerical stability during training \cite{liu2022swin}. Moreover, the architecture employs scaled cosine attention instead of dot product between queries and keys. This approach effectively reduces the dominance of some attention heads for a few pixel pairs, leading to better overall performance.
Inspire by \cite{lu2021transformer}, we propose an RS2TB apply it in image compression, as shown in Fig.\ref{FIG:2}. Similar to the STB \cite{lu2021transformer}, the RS2TB utilizes feature embedding(FE) and feature unembedding(FU) to change the dimension of input image. The first feature embedding (FE) layer projects input features with the size of $H\times W \times C$ into the dimension of $HW\times C$, followed by SwinV2 Attention, the normalization layer (LN). And the MLP layer together form SwinV2 transformer for easy calculation of window-based self-attention, with the final FU layer remapping the attention-weighted features back to their original size $H \times W \times C$. Furthermore, we incorporate Skipping joins into our architecture to enhance feature aggregation, resulting in improved compression performance.

\begin{figure*}
\centering
\subfigure{
\begin{minipage}[ht]{0.5\linewidth}
\centering
\includegraphics[width=\columnwidth]{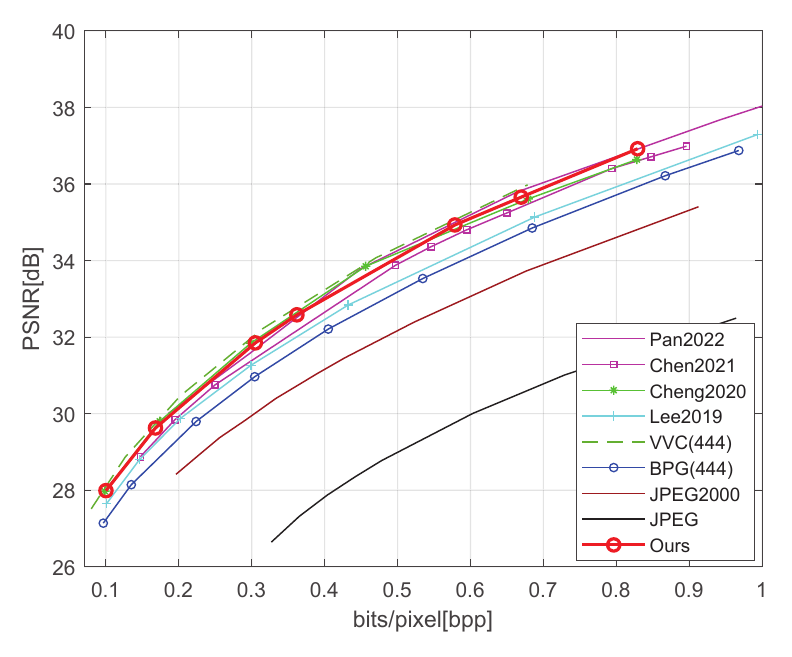}
\end{minipage}
}%
\subfigure{
\begin{minipage}[ht]{0.5\linewidth}
\centering
\includegraphics[width=\columnwidth]{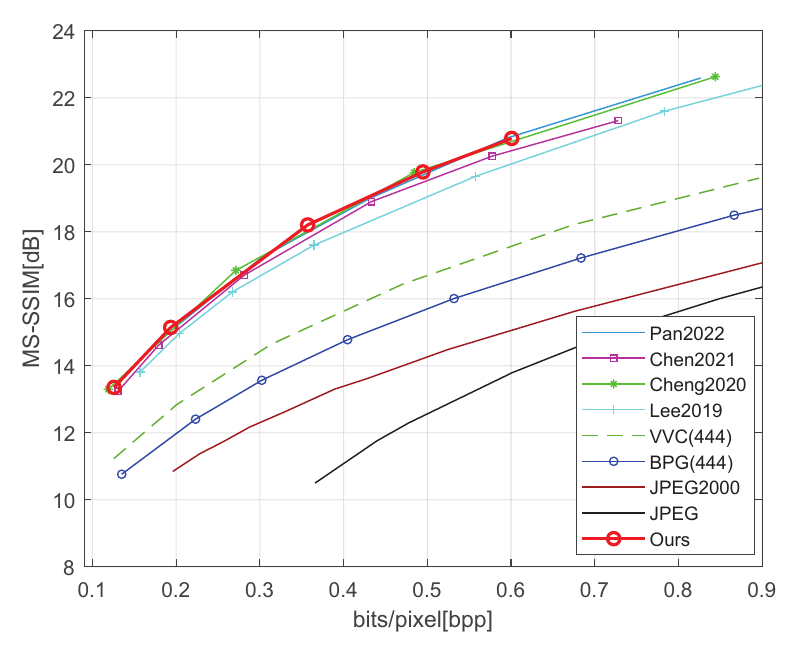}
\end{minipage}
}%
\centering
\caption{Average comparison results on all 24 Kodak images in terms of PSNR and MS-SSIM.}
\label{test_Kodakdataset}
\end{figure*}

\begin{figure*}
\centering
\subfigure{
\begin{minipage}[ht]{0.5\linewidth}
\centering
\includegraphics[width=\columnwidth]{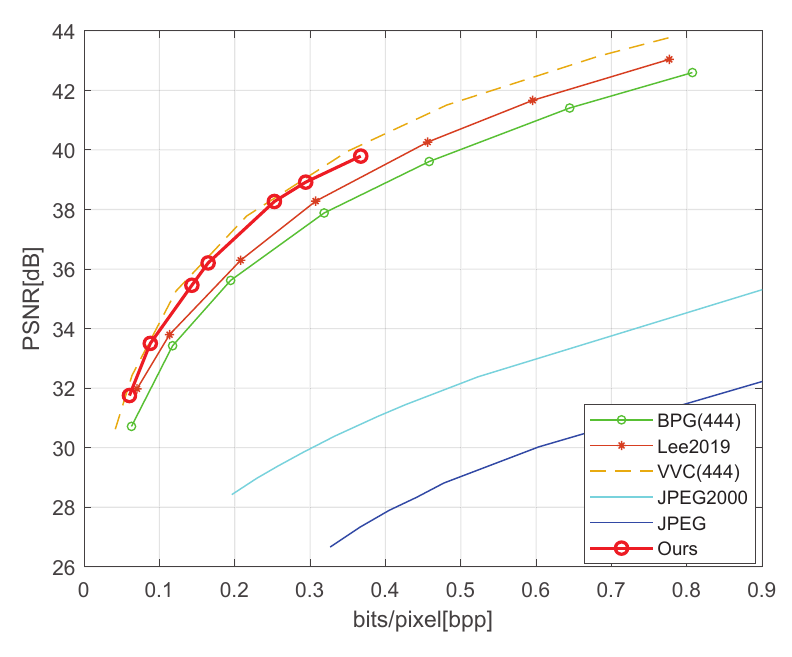}
\end{minipage}
}%
\subfigure{
\begin{minipage}[ht]{0.5\linewidth}
\centering
\includegraphics[width=\columnwidth]{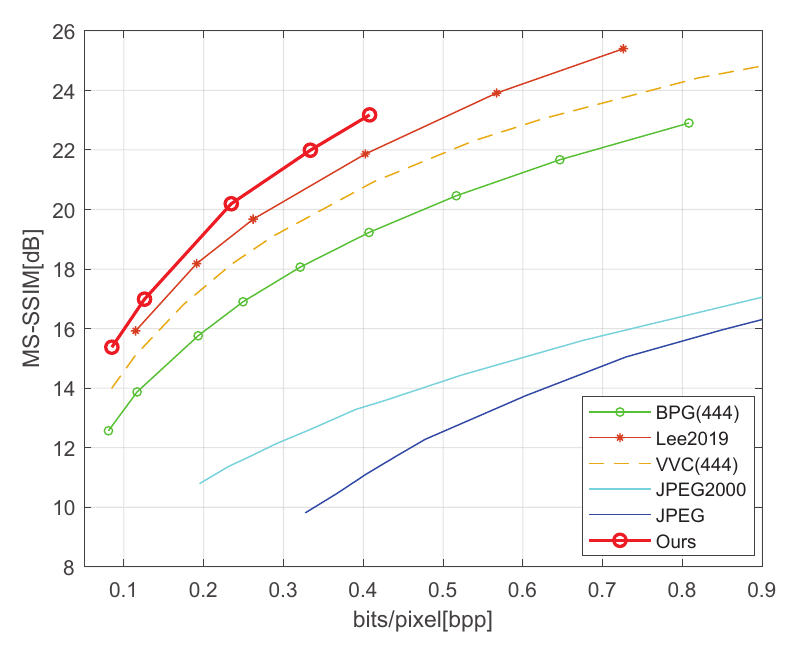}
\end{minipage}
}%
\centering
\caption{Average comparison results on all 40 Tecnick images in terms of PSNR and MS-SSIM.}
\label{test_Tecnickdataset}
\end{figure*}

\subsection{Loss Function}
In the encoding part, the input image x is encoded into latent features y, and then y is quantized into $\hat{y}$, which is decoded back to the reconstructed image $\hat{x}$ in the decoder. In order to obtain different bit rates, we trained several independent models with different lagrange multiplier $\lambda$ values. The optimization objective is to minimize the rate-distortion cost through an end-to-end learning means:
 \begin{equation}\label{lossfun}
 L = R(\hat{y}) + \lambda D(x, \hat{x})
 \end{equation} 

where R is compressed bit rate of $\hat{y}$ and the distortion D between the ground truth x and reconstruction $\hat{y}$. The distribution of the rate R is the entropy $\hat{y}$, which is estimated by a entropy model $p_{\hat{y}|\theta}$ during the training. The specific equation is shown as follows:

 \begin{equation}\label{entropy model}
 \text{R} = \text{E}[-\log_2{p_{\hat{y}|\theta}}({\hat{y}|\theta})]
 \end{equation} 

The distortion D is defined as D = MSE(x, $\hat{x}$) for MSE optimization and D = 1 - MS-SSIM(x, $\hat{x}$) for MS-SSIM \cite{ms-ssim} optimization. We adapt $\lambda$ for rate-distortion trade-off at various bit rates.

\section{Experimental Results}
\subsection{Experiment settings}
 \textbf{ Training Details:} Following the previous works, we use the Flicker 2W \cite{flicker2w} for training. This dataset is built for evaluating different image processing tasks, which contains 20745 high-quality general images. During the training, we randomly crop these images into fixed patch at a size of $256 \times 256\times3$. The proposed model is implemented on the open-source CompressAI PyToych library \cite{compressai}. All the experiments are conducted on RTX 2080 Ti GPU and trained for 400 epochs with the learning rate of 10-4.  And the Adam \cite{adam} is used as the optimizer for the whole training.
 
We use the mean squared error(MSE) and MS-SSIM as the quality metric to optimize our models. For the MSE metric, the parameter $\lambda$ is chosen from the set \{0.0016, 0.0032, 0.0075, 0.015, 0.023, 0.03, 0.045\}, The number of channels N in the latent representation is set to 128 for the first four cases for lower-rate models, and is increased to 192 for the last three cases for higher-rate models. When the MS-SSIM metric is used, the parameter $\lambda$  is set to \{6,12,40,80,120\}. The value of N is set to 128 for the first two cases, and 192 for the other three cases. Other parameters follow the setting in \cite{lu2021transformer}. We use RTX 2080Ti and 2.9GHz Intel Xeon Gold 6226R CPU to complete the following experiments.

\textbf{Evaluation:} The test datasets are the Kodak dataset \cite{Kodak_dataset} and Tecnick dataset \cite{Tecnick_dataset}. The Kodak dataset consists of 24 images with resolution of 768x512 or $512 \times 768$. The Tecnick dataset contains 40 images with high resolutions of $1200 \times 1200$. We evaluate our model with the the peak signal-to-noise ratio (PSNR) and the multiscale structural similarity index (MS-SSIM) \cite{ms-ssim} to quantify the image quality and the bits per pixel (bpp) to measure the bit rate.   

According to different evaluation results, we draw the rate-distortion curves. 

\subsection{Complexity Comparison}
To compare the complexity of the model size, we select the several known as models, including Lee2019\cite{lee2018context}, Cheng2020 \cite{cheng2020learned}, Chen2021 \cite{chen2021} and the traditional most advanced codec VVC \cite{VVC}. In order to better observe and compare the codec time and model complexity of these models, we conducted evaluation tests on Kodak dataset and Tecnick dataset respectively. The number filter is set to 128 for low bits and 192 for high bits. After completing  the evaluation, average values are calculated from the low and high bit model results respectively for comparison. Table \ref{tab1} shows the results of the complexity comparison of several different models. 

As shown in Table \ref{tab1}, the encoding time of VVC on both Kodak datasets and Tecnick datasets are the longest. However, once the encoding is complete, the decoding speed is very fast, with the average decoding time of 0.73s at low bit rate and about 1.3s at high bit rate. Compared to Cheng et al.'s \cite{cheng2020learned} method, our method shows similar performance at low bit rates but with significantly fewer model size, only 57.09 \% of Cheng et al.'s \cite{cheng2020learned} method. This indicates that our method can achieve similar performance with less computing resources and storage space. At high bit rates, our method demonstrates faster codec time, and the number of model size is only 56.81\% of Cheng et al.'s \cite{cheng2020learned} method. Moreover, our approach outperforms Cheng et al.'s \cite{cheng2020learned} in image compression. At the same bit rate, our method can achieve higher compression quality due to the better utilization of spatial correlation in the image and the use of more suitable visual quality evaluation indicators in the loss function. Overall, our method provides a more efficient and effective solution for image compression than Cheng2020 \cite{cheng2020learned}. 

\begin{table*}
\caption{The compare of the Encoding, Decoding time and Model size on Kodak and Tecnick datasets.}
\begin{center}
  \begin{tabular}{cccccccc}
  \hline
  \multirow{2}{*}{\textbf{dataset}} & \multirow{2}{*}{\textbf{Method}} & \textbf{}& \textbf{Low Bit Rate} & \textbf{} & \textbf{}& \textbf{High Bit Rate} & \textbf{}\\
  \textbf{}&\textbf{}  & \textbf{Enc Time(s)} & \textbf{Dec Time(s)}& \textbf{Model size} & \textbf{Enc Time(s)} & \textbf{Dec Time(s)}& \textbf{Model size}\\
  \hline
  \multirow{5}{*}{Kodak} & \text{VVC} & 402.27 & 0.60 & None & 760.81 & 0.81 & None   \\
  \textbf{}& \text{Lee2019 \cite{lee2018context}} & 10.38 & 38.25 & 123.8MB & 21.65 & 71.44 & 292.6MB \\
  \textbf{}& \text{Cheng2020\cite{cheng2020learned}} & 20.43   & 23.04 & 57.8MB & 91.05 & 93.28 & 126.9MB  \\
  \textbf{}& \text{Chen2021\cite{chen2021}}  & 400.26  & 2315.07 & 300.9MB & 365.18 & 8415.14 & 300.9MB  \\
  \textbf{}& \textbf{Ours}  & \textbf{23.24 } & \textbf{31.83} & \textbf{33.0MB} & \textbf{31.83} & \textbf{39.66} & \textbf{72.1MB}  \\
  \hline
  \multirow{5}{*}{Tecnick} & \text{VVC} & 235.46 & 0.875 & None & 2156.59 & 1.794 & None   \\
  \textbf{}& \text{Lee2019 \cite{lee2018context}} & 58.18 & 148.05 & 123.8MB & 110.42 & 291.60 & 292.6MB \\
  \textbf{}& \text{Cheng2020\cite{cheng2020learned}} & 51.28  & 53.88 & 57.8MB & 398.96 & 414.32 & 126.9MB  \\
  \textbf{}& \text{Chen2021\cite{chen2021}}  & 65.43  & 4503.25 & 300.9MB & 167.12 & 4983.6 & 300.9MB  \\
  \textbf{}& \textbf{Ours}  & \textbf{87.18} & \textbf{116.90} & \textbf{33.0MB} & \textbf{127.36} & \textbf{157.25} & \textbf{72.1MB}  \\
  \hline
\end{tabular}
\label{tab1}
\end{center}
\end{table*}

\begin{figure}
		\includegraphics[scale=0.6]{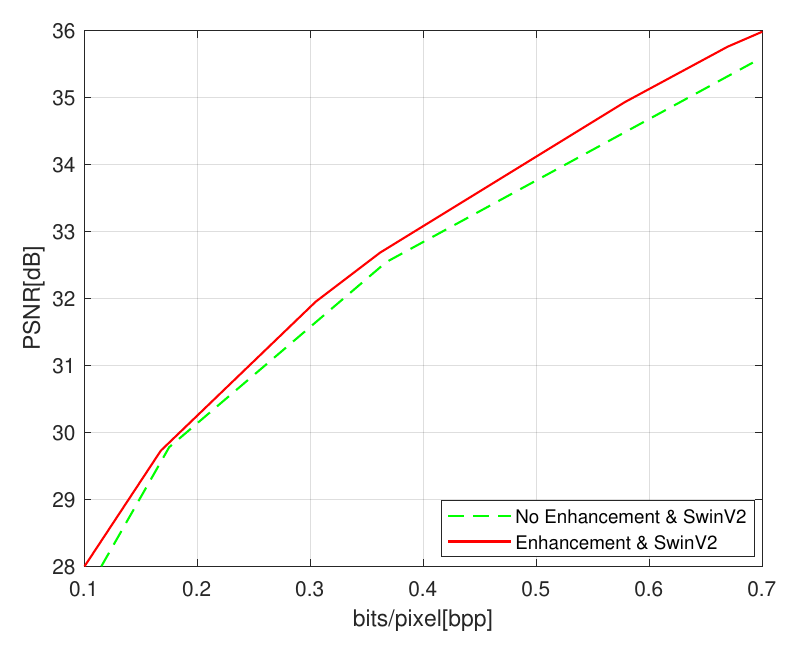}
	\caption{The performance comparison of different components on kodak datasets. }
	\label{FIG:5}
\end{figure}

\subsection{Rate-distortion Performance}
In this section, we compared our model with some other learning-based models, including \cite{Pan_Lu_Hu_Xu_2022}, \cite{chen2021}, \cite{cheng2020learned} and \cite{lee2018context}. The traditional image compression codecs, including VVC \cite{VVC}, BPG \cite{Bpg_125}, JPEG2000 and JPEG in terms of both PSNR and MS-SSIM metrics. To enhance clarity, MS-SSIM values are converted to $-10\log_{10}(1-{MS-SSIM})$ for better comparison. 

The rate-distortion curves on Kodak dataset is shown in Fig.\ref{test_Kodakdataset}. When optimized for PSNR, our method almost achieves the same performance with Cheng2020\cite{cheng2020learned} at low bit rate, and outperforms Cheng2020\cite{cheng2020learned} at high bit rates. With a PSNR improvement of 0.23dB at 0.8bpp, which is almost on par with VVC. The model parameters used are only half of Cheng2020\cite{cheng2020learned}. When we use MS-SSIM to optimize the model, the results shows that our model performed almost as well as Cheng2020\cite{cheng2020learned} and outperformed the models from Lee2019\cite{lee2018context}, Chen2021\cite{chen2021} and Pan2022\cite{Pan_Lu_Hu_Xu_2022}. It is worth noting that optimizing with MS-SSIM resulted in a 3.48 dB improvement compared to optimizing with PSNR, and the latter is better than VVC.

Fig.\ref{test_Tecnickdataset} shows the R-D performances on Tecnick dataset. The Tecnick dataset consists mainly of high-resolution images. We choose three traditional compression methods and another learning-based compression method \cite{lee2018context}. Our model achieves almost the same performance as VVC and is better than Lee2019\cite{lee2018context} on PSNR. When optimized with MS-SSIM, our performance is significantly better than that of several other models compared.

\begin{figure*}
\centering
\subfigure[Original]{
\begin{minipage}[t]{0.33\linewidth}
\centering
\includegraphics[scale=0.21]{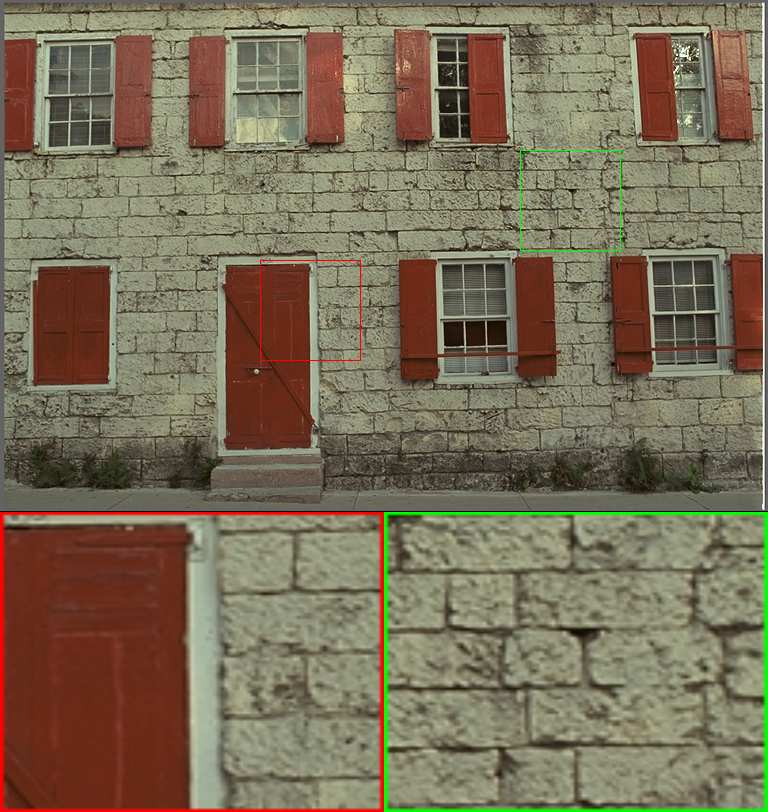}
\end{minipage}
}%
\subfigure[JPEG(0.191/19.95/0.700)]{
\begin{minipage}[t]{0.33\linewidth}
\centering
\includegraphics[scale=0.21]{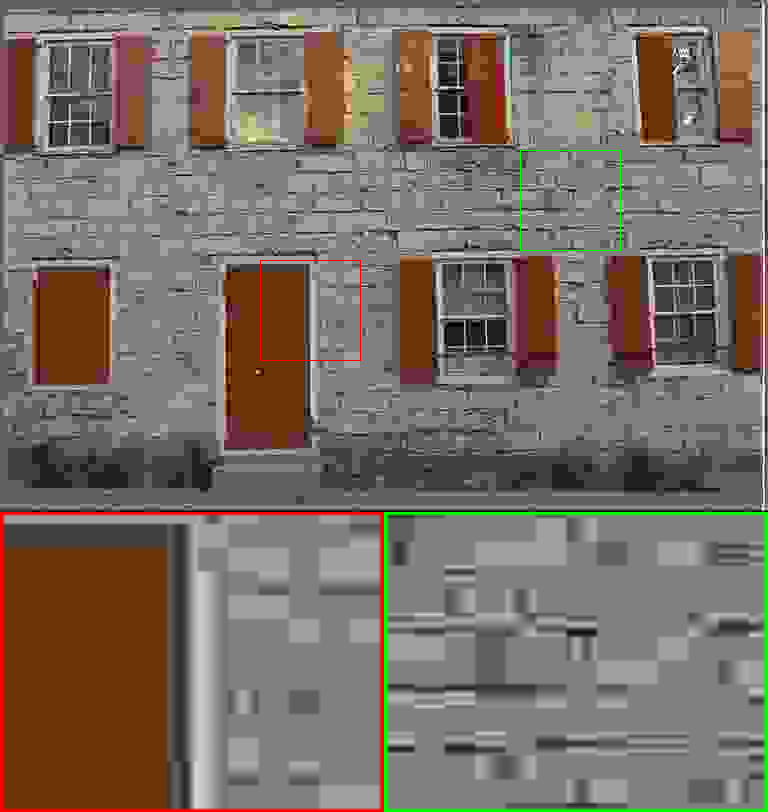}
\end{minipage}
}%
\subfigure[BPG(0.157/25.00/0.897)]{
\begin{minipage}[t]{0.33\linewidth}
\centering
\includegraphics[scale=0.21]{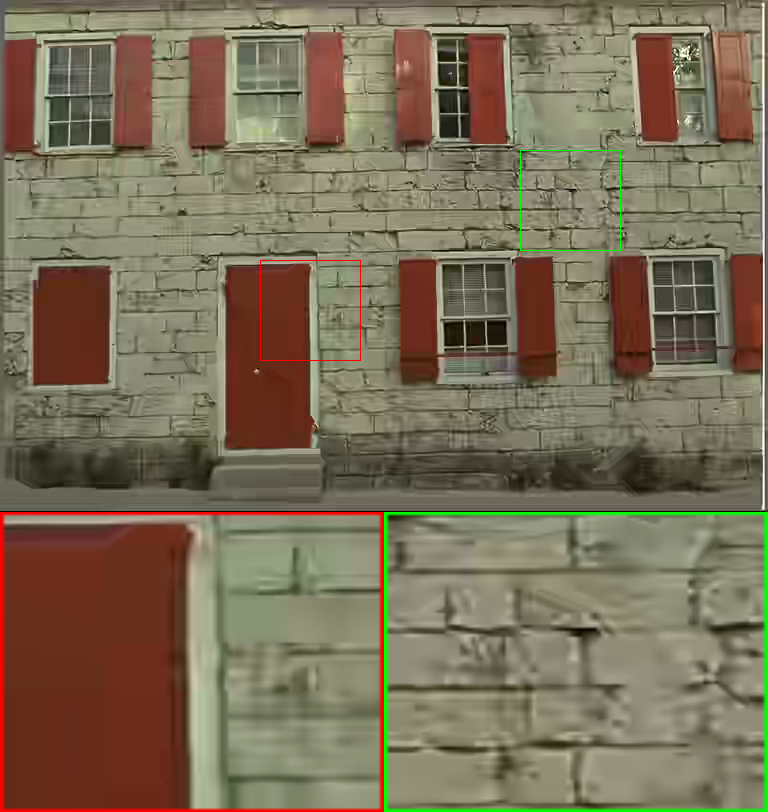}
\end{minipage}
}%

\subfigure[VVC(0.160/25.83/0.915)]{
\begin{minipage}[t]{0.33\linewidth}
\centering
\includegraphics[scale=0.21]{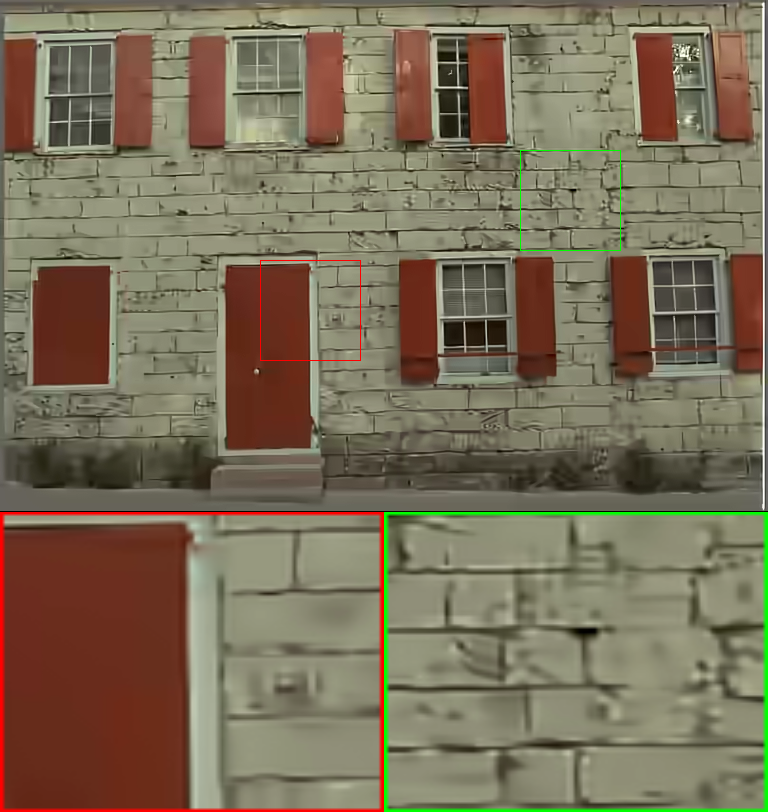}
\end{minipage}
}%
\subfigure[Ours(0.138/25.26/0.909)]{
\begin{minipage}[t]{0.33\linewidth}
\centering
\includegraphics[scale=0.21]{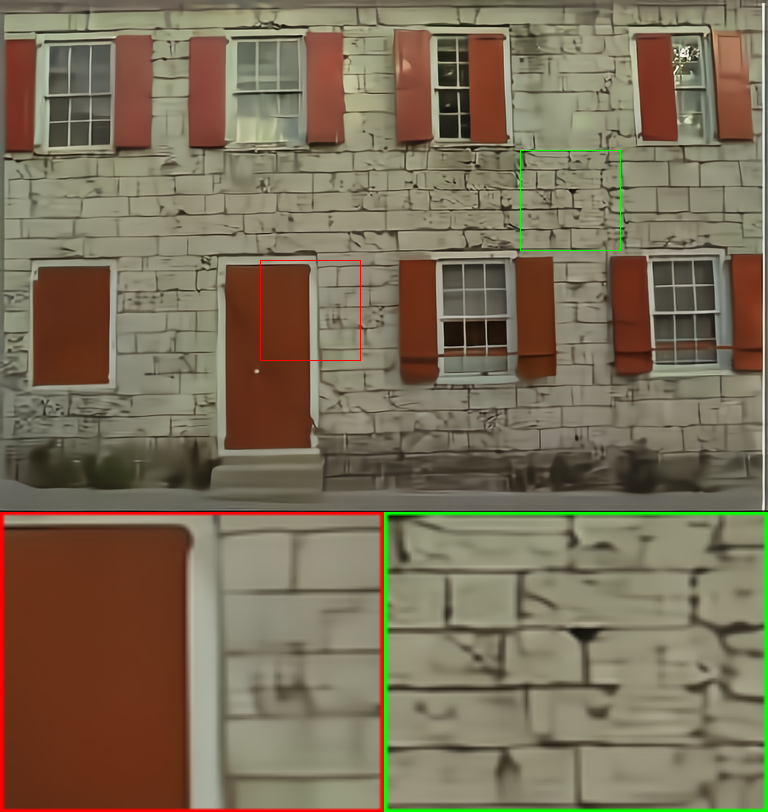}
\end{minipage}
}%
\subfigure[MS-SSIM(0.181/24.23/0.946)]{
\begin{minipage}[t]{0.33\linewidth}
\centering
\includegraphics[scale=0.21]{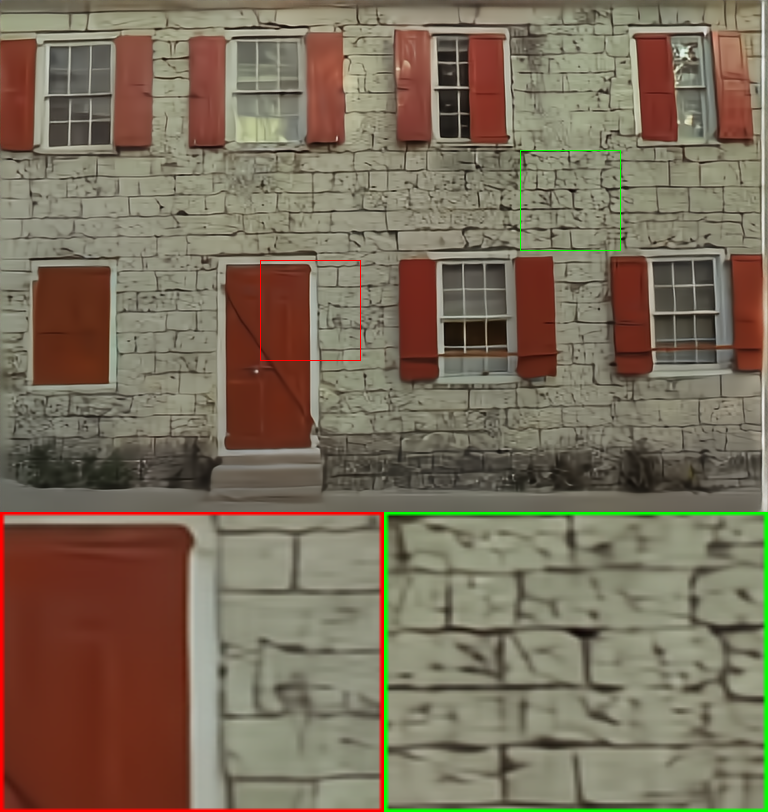}
\end{minipage}
}%
\centering
\caption{Example origin01 in the Kodak dataset (bpp, PSNR(dB), MS-SSIM).}
\label{Example_1}
\end{figure*}

\begin{figure*}
\centering
\subfigure[Original]{
\begin{minipage}[t]{0.33\linewidth}
\centering
\includegraphics[scale=0.21]{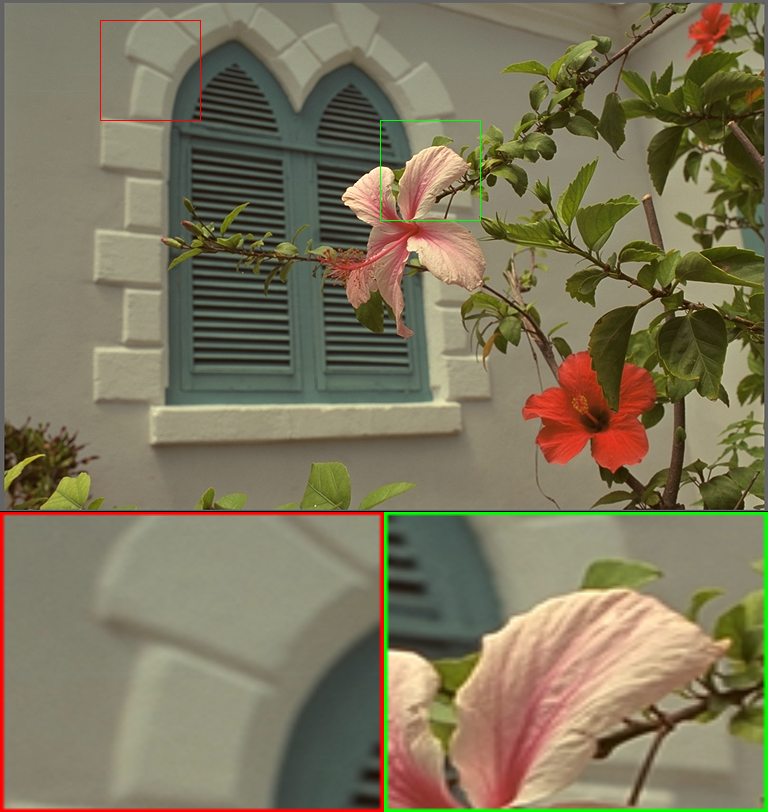}
\end{minipage}
}%
\subfigure[JPEG(0.171/21.88/0.793)]{
\begin{minipage}[t]{0.33\linewidth}
\centering
\includegraphics[scale=0.21]{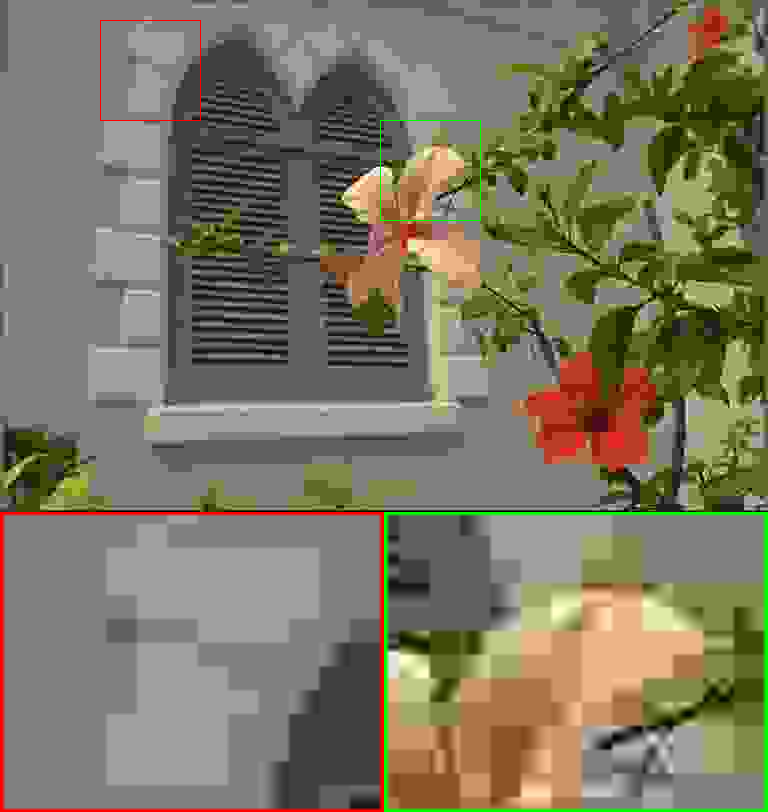}
\end{minipage}
}%
\subfigure[BPG(0.091/28.21/0.938)]{
\begin{minipage}[t]{0.33\linewidth}
\centering
\includegraphics[scale=0.21]{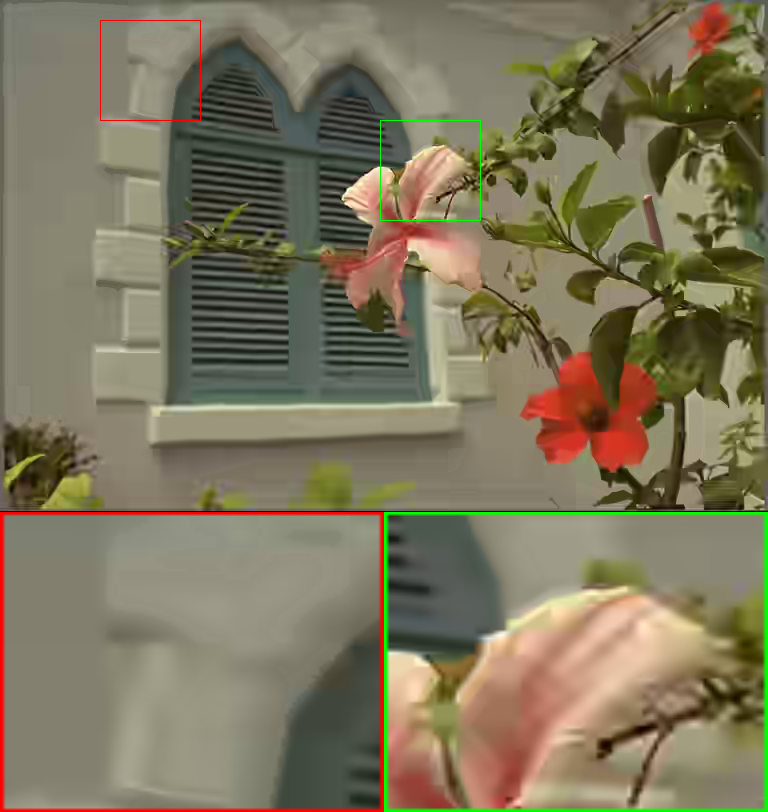}
\end{minipage}
}%

\subfigure[VVC(0.092/29.58/0.953)]{
\begin{minipage}[t]{0.33\linewidth}
\centering
\includegraphics[scale=0.21]{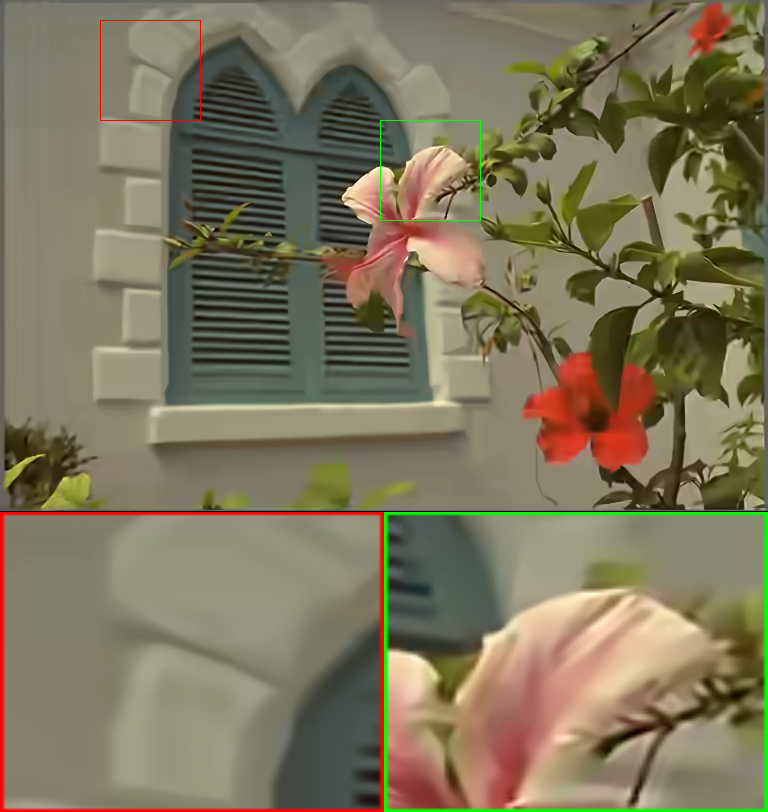}
\end{minipage}
}%
\subfigure[Ours(0.090/29.49/0.957)]{
\begin{minipage}[t]{0.33\linewidth}
\centering
\includegraphics[scale=0.21]{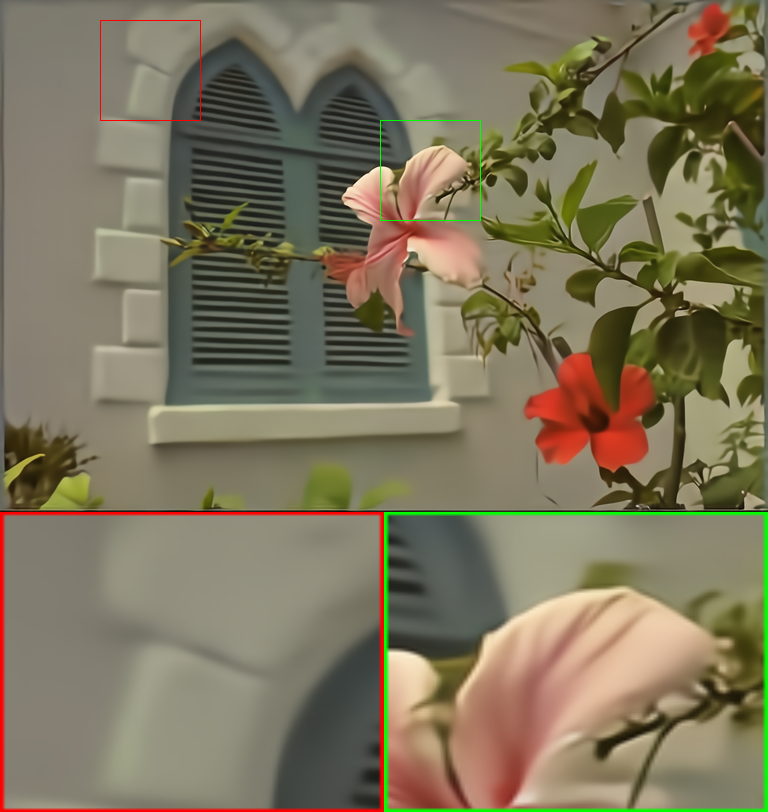}
\end{minipage}
}%
\subfigure[MS-SSIM(0.105/27.77/0.973)]{
\begin{minipage}[t]{0.33\linewidth}
\centering
\includegraphics[scale=0.21]{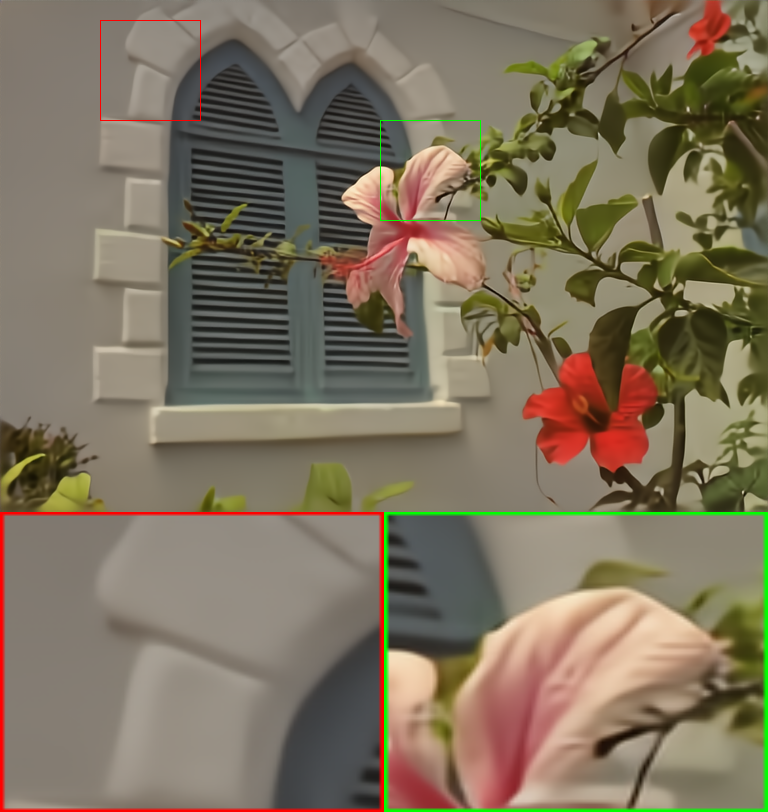}
\end{minipage}
}%
\centering
\caption{Example origin07 in the Kodak dataset (bpp, PSNR(dB), MS-SSIM).}
\label{Example_2}
\end{figure*}

\begin{figure*}
\centering
\subfigure[Original]{
\begin{minipage}[t]{0.33\linewidth}
\centering
\includegraphics[scale=0.21]{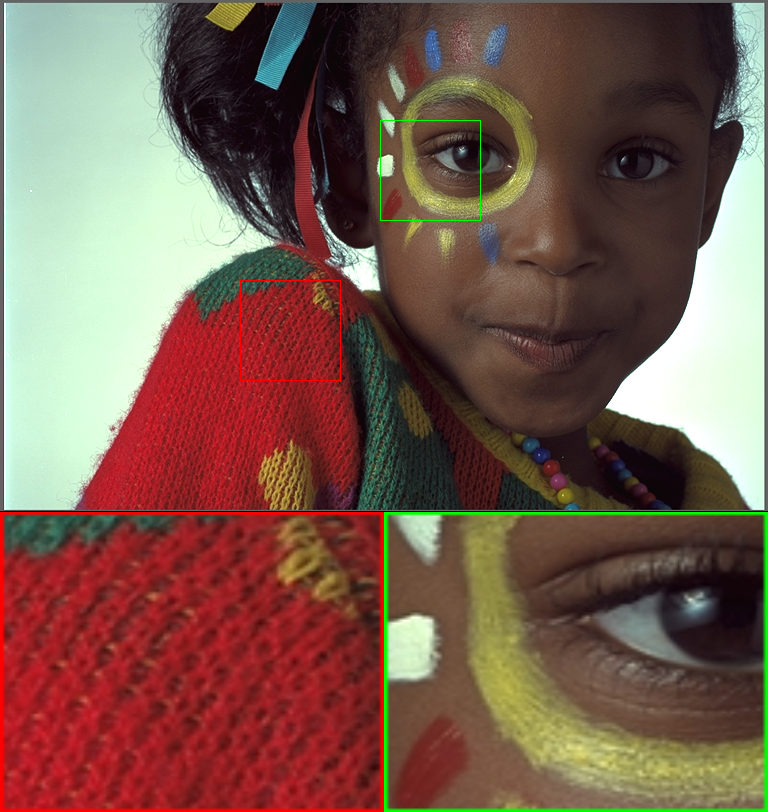}
\end{minipage}
}%
\subfigure[JPEG(0.166/21.82/0.730)]{
\begin{minipage}[t]{0.33\linewidth}
\centering
\includegraphics[scale=0.21]{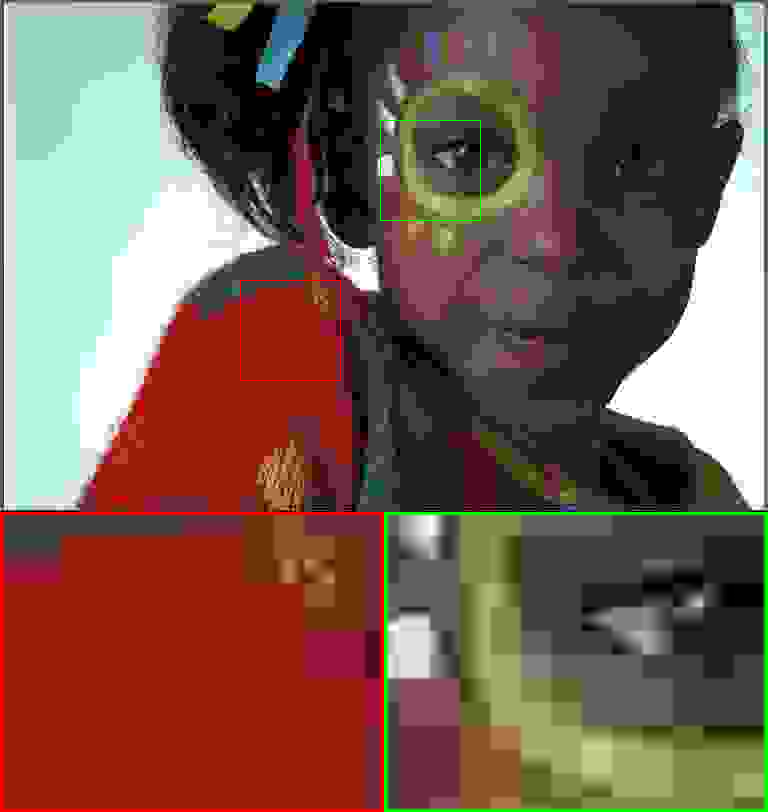}
\end{minipage}
}%
\subfigure[BPG(0.053/28.61/0.729)]{
\begin{minipage}[t]{0.33\linewidth}
\centering
\includegraphics[scale=0.21]{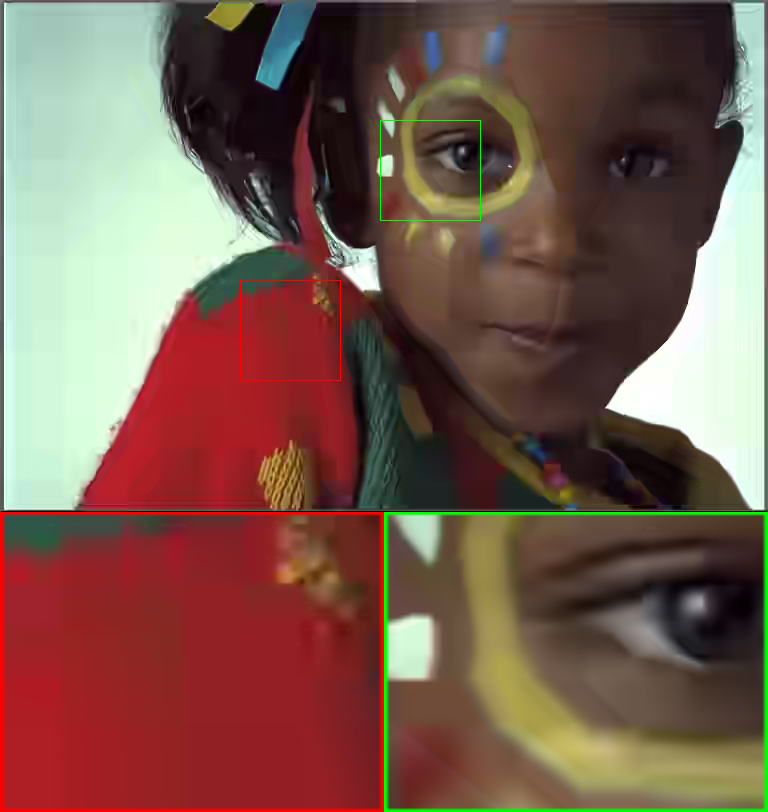}
\end{minipage}
}%

\subfigure[VVC(0.052/29.63/0.923)]{
\begin{minipage}[t]{0.33\linewidth}
\centering
\includegraphics[scale=0.21]{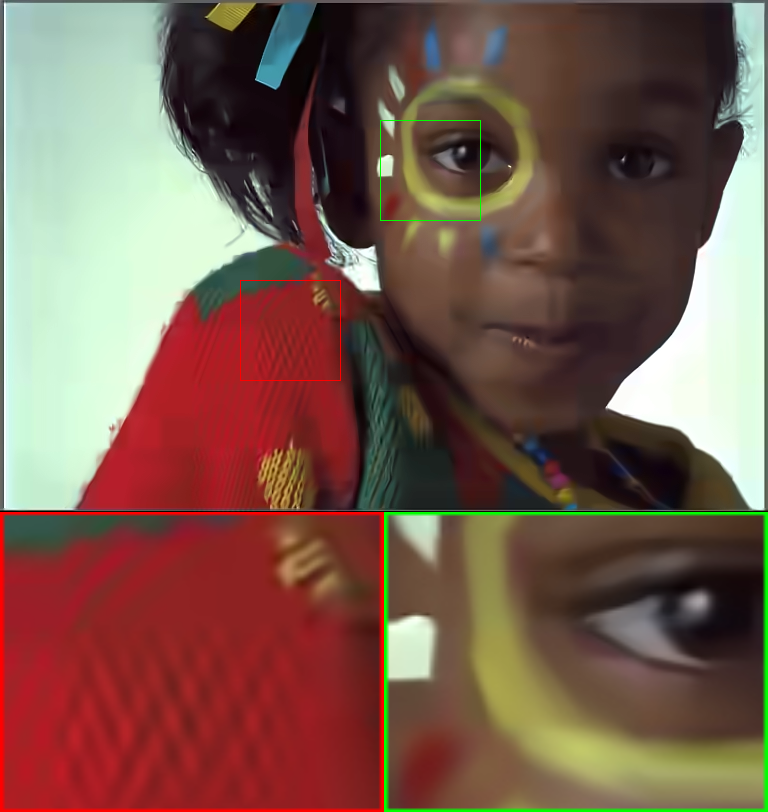}
\end{minipage}
}%
\subfigure[Ours(0.056/29.24/0.930)]{
\begin{minipage}[t]{0.33\linewidth}
\centering
\includegraphics[scale=0.21]{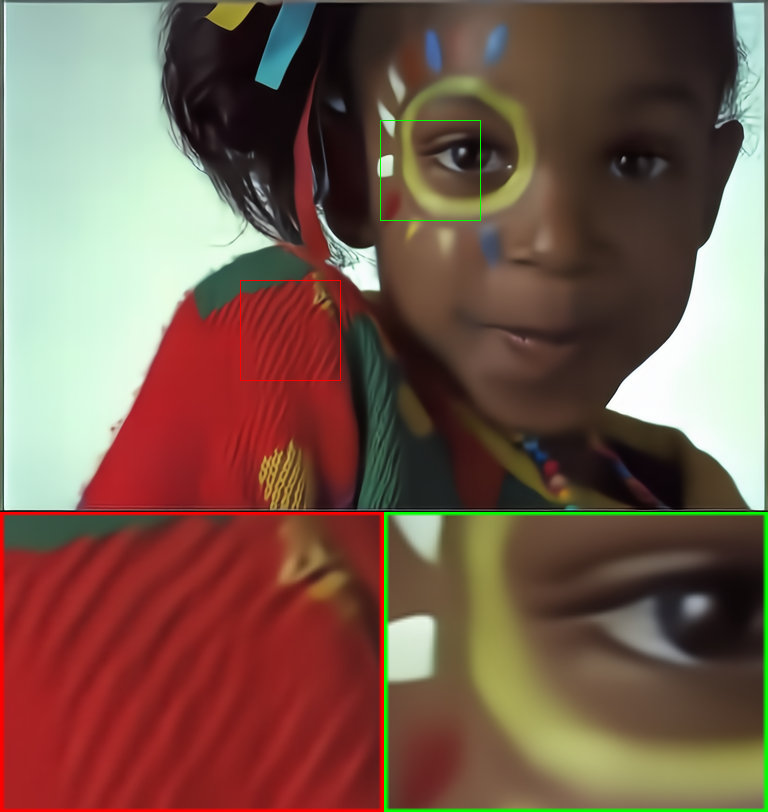}
\end{minipage}
}%
\subfigure[MS-SSIM(0.079/26.42/0.961)]{
\begin{minipage}[t]{0.33\linewidth}
\centering
\includegraphics[scale=0.21]{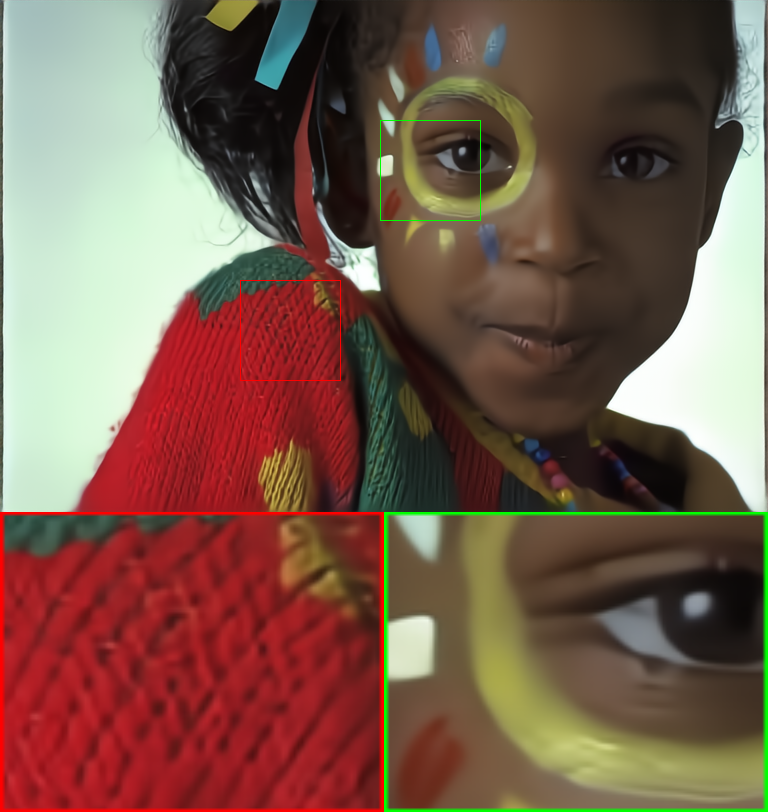}
\end{minipage}
}%
\centering
\caption{Example origin15 in the Kodak dataset (bpp, PSNR(dB), MS-SSIM).}
\label{Example_3}
\end{figure*}

In addition, we show three examples in Fig.\ref{Example_1}, Fig.\ref{Example_2} and Fig.\ref{Example_3} to compare visual quality, our model achieves the best visual effect in different ways.

\begin{table*}
\caption{The detailed comparison of different components.}
\begin{center}
  \begin{tabular}{cccccccc}
  \hline
  \textbf{Method}&\textbf{Number Filters}  & \textbf{$\lambda$} & \textbf{Bpp}& \textbf{PSNR} & \textbf{MS-SSIM} & \textbf{Encoding Time} & \textbf{Decoding Time}\\
  \hline
  \textbf{SwinV2}           & 128 & 0.0032 & 0.175 & 29.78dB & 0.944 & 17.91s & 24.86s   \\
  \textbf{Enhance+SwinV2}   & 128 & 0.0032 & 0.168 & 29.73dB & 0.942 & 24.30s & 33.48s \\
  \textbf{SwinV2}           & 192 & 0.03   & 0.692 & 35.32dB & 0.984 & 27.09s & 34.30s  \\
  \textbf{Enhance+SwinV2}   & 192 & 0.03   & 0.670 & 35.66dB & 0.985 & 30.98s & 39.05s  \\
  \hline
\end{tabular}
\label{tab2}
\end{center}
\end{table*}

\subsection{Ablation Studies}

 In order to compare with different components and further verify the influence of feature enhancement module on performance, corresponding ablation experiments are conducted. Similar to the previous experiment, we trained 400 epochs on the Filker 2W \cite{flicker2w} dataset using different bit rates. As shown in the Fig.\ref{FIG:5}, we can observe that SwinV2 transformer can effectively improve compression performance after non-linear feature enhancement module. Detailed data on the compression performance of the different components is shown in Table \ref{tab2}

 \subsection{Qualitative Results}
 We select three examples in Fig.\ref{Example_1}, Fig.\ref{Example_2} and Fig.\ref{Example_3} for qualitative comparison of visual visualization. We select image origin 01, image origin 07, and image origin 15 from the Kodak dataset as samples for our evaluation.  To facilitate detailed observation and comparison, we choose the lowest bit rate during the comparison. It can be seen that our model achieves almost the same performance as VVC when adopting MSE optimization. Far better than JPEG and BPG encoder performance. When optimizing for MS-SSIM, our method preserves more details in the reconstructed image, making it visually more similar to the original image. 

\section{Conclusion}

In this paper, we propose an enhanced residual SwinV2 transformer for learned image compression framework. It can achieve better performance than Cheng with nearly half the model size saved. Meanwhile, we introduce improvements to the transformer network, including non-linear feature enhancement before the convolution operation. Our performance in PSNR and MS-SSIM metric is better than that of BPG and other learning-based image compression methods. And in high resolution pictures our model achieved the best performance in MS-SSIM. In the future work, we will continue to explore the improvement of the image coding process, enhance the ability to extract global information from the model, and further reduce the complexity of the model to achieve better compression performance.

\bibliographystyle{cas-model2-names}

\bibliography{cas-refs}

\end{document}